\documentclass[aps,prd,preprint,groupedaddress]{revtex4-1}

 \usepackage{graphicx} 

\begin{document}

\title{  Holographic model for charmonium dissociation   }



\author{Nelson R. F. Braga}\email{braga@if.ufrj.br}
\affiliation{Instituto de F\'{\i}sica,
Universidade Federal do Rio de Janeiro, Caixa Postal 68528, RJ
21941-972 -- Brazil}

\author{Luiz F.  Ferreira}
\email{luizfaulhaber@if.ufrj.br}
\affiliation{Instituto de F\'{\i}sica,
Universidade Federal do Rio de Janeiro, Caixa Postal 68528, RJ
21941-972 -- Brazil}

\author{ Alfredo Vega}
\email{alfredo.vega@uv.cl}
\affiliation{Instituto de F\'{\i}sica y Astronom\'{\i}a, Universidad de Valpara\'{\i}so,
Avenida Gran Bretana 1111, Valpara\'{\i}so, Chile}


\begin{abstract}
 We present a holographic bottom up model for the thermal behavior  of $ c \bar c$  vector mesons  in a finite temperature and density plasma.  There is a clear physical interpretation for the three input energy parameters of the model. Two of them are related to the mass spectrum of the heavy meson. Namely the quark mass and the string tension of the quark anti-quark interaction.
The third parameter is a large energy scale associated with the non-hadronic meson decay. In such a process the heavy meson is transformed into a much lighter state by  electroweak processes. The corresponding transition amplitude is assumed to depend on the enegy scale associated with this large mass variation. With this  three parameter model one can fit the masses and decay constants of $ J / \Psi $ 
 and three radial excitations with an rms error of 20.7\%.
 Using the geometry of a charged black hole, one finds the spectral function for charmonium states inside a plasma at finite temperature and density.  The charmonium  dissociation in the medium 
is represented by the decrease in the height of the spectral function peaks.  
\end{abstract}

\keywords{Gauge-gravity correspondence, Phenomenological Models}

\maketitle

\section{ Introduction }   

Heavy vector mesons, detected after a heavy ion collision, can provide information about the possible existence of a plasma state for very short time scales. The suppression of such particles could indicate their dissociation in this thermal medium\cite{Matsui:1986dk} (see also \cite{Satz:2005hx}). It is, thus,  important to understand the thermal properties, in particular the dissociation temperatures,  of heavy mesons. 

A bottom up holographic model for $ b \bar b$  vector mesons (bottomonium)  in a plasma at finite temperature but  zero density appeared recently in ref. \cite{Braga:2016wkm}.  An extension to finite density was then developed in \cite{Braga:2017oqw}. 
The first radial excitations $1S, 2S$ and $ 3S$ appear as clear peaks of the spectral function. The height of the peaks  decrease as the temperature of the medium increases, representing the dissociation of the  states.  The holographic model used in \cite{Braga:2016wkm,Braga:2017oqw} appeared initially in \cite{Braga:2015jca}  for the zero temperature (vacuum)  case. 
 This model was built up with the purpose of  overcoming a problem of the so called AdS/QCD approach.
 Namely experimental data show that decay constants decrease monotonically with radial excitation level.
 This fact was contrasting with results of previous holographic models.  
 It was shown in ref. \cite{Braga:2015jca} that the addition of an ultraviolet (UV) energy parameter in the soft wall model, associated with the non hadronic decay of mesons, leads to decay constants decreasing with excitation level,
 The model includes two energy parameters: the infrared mass scale of the soft wall and the UV scale that is the inverse of the position of anti-de Sitter space where correlation functions are calculated. 

Reference  \cite{Braga:2016wkm} includes also a study of charmonium states. But the holographic picture obtained for   $ J / \Psi $  at finite temperature is not appropriate:  the dissociation takes place  at too small temperatures. A possible explanation for the failure to describe charmonium is that the decay constant obtained for $ J / \Psi $  is 40\% smaller than the experimental result. This contrasts with bottomonium  $\Upsilon$ state, for which the model of \cite{Braga:2016wkm,Braga:2015jca} provides a decay constant just 12\% smaller than the experimental result. 

The connection   between decay constants and finite temperature behavior is a consequence of the 
relation between these quantities and the spectral function.  
The thermal spectral function is the imaginary part of the retarded Green's function. 
The relevant part of the Green function is the two point function that, at zero temperature, has a spectral decomposition  in terms of masses $m_n$ and   decay constants $f_n$  of the states:  
 \begin{equation}
\Pi (p^2)  = \sum_{n=1}^\infty \, \frac{f_n^ 2}{(- p^ 2) - m_n^ 2 + i \epsilon} \,.
\label{2point}
\end{equation} 
The imaginary part of eq.(\ref{2point}) is a sum of delta peaks with coefficients proportional to the square of the decay constants: $ f_n^2 \, \delta ( - p^2 - m_n^2 ) $. 
At finite temperature, the quasi-particle states appear in the spectral function as smeared peaks that decrease as the temperature $T$ and/or the density $\mu$ of the medium increase. This analysis suggests that a consistent  extension of a hadronic model to finite temperature should take into account the zero temperature behavior, where decay constants play an essential role. 
    
The holographic model for heavy vector meson of ref. \cite{Braga:2015jca} 
 includes one parameter associated with the mass spectrum and one associated with the decay process. 
A limitation of this model is that actually the mass spectrum of a heavy meson depends itself on two parameters:  the quark mass, that is not negligible in this case, and the string tension that represents the strong interaction between quarks. 
A phenomenological model designed for masses and decay constants should include three parameters. The third one associated with the non hadronic decay. In such processes the heavy meson transforms into much lighter particles.
It is reasonable to assume that the  transition amplitude depends of the mass difference between the initial and final states, that is of the order of the heavy meson mass. 

Another aspect of reference \cite{Braga:2015jca}  is that there is a ultraviolet hard wall:  a finite position of AdS space where the solutions describing states vanish. 
The purpose of this letter is to develop a holographic model for charmonium with three parameters, with the roles discussed above: quark mass, string tension and mass transition scale for non hadronic decay. The new framework developed here contains just a smooth background, without any hard wall.  The 
 finite temperature and density picture that emerges from this new model provides a consistent description for the dissociation of charmonium states.  This letter is organized as follows. In section II we describe the calculation of masses and decay constants in soft wall like models and present the new model at zero temperature. The extension to finite temperature and density is shown in section III and the procedure for the calculation of the spectral function in section IV.
 The results and discussions appear in section V.

\section{Holographic model for charmonium masses and decay constants}

Let us consider a generalized version of the soft wall model\cite{Karch:2006pv}. Vector mesons are described by a vector  field $V_m = (V_\mu,V_z)\,$ ($\mu = 0,1,2,3$), which is dual to the gauge theory current $ J^\mu = \bar{\psi}\gamma^\mu \psi \,$. The action is:
\begin{equation}
I \,=\, \int d^4x dz \, \sqrt{-g} \,\, e^{- \phi (z)  } \, \left\{  - \frac{1}{4 g_5^2} F_{mn} F^{mn}
\,  \right\} \,\,, 
\label{vectorfieldaction}
\end{equation}
where $F_{mn} = \partial_m V_n - \partial_n V_m$ and $\phi(z)$ is a background  dilaton field. The space is anti-de Sitter with metric 
\begin{equation}
 ds^2 \,\,= \,\, e^{2A(z)}(-dt^2 + d\vec{x}\cdot d\vec{x} + dz^2)\,,
\end{equation}
\noindent where $ A(z) = -log(z/R) $.  Choosing the gauge $V_z=0$  the equation of motion for the transverse (1,2,3) components  of the  field, denoted generically as $V$, in momentum space reads
\begin{equation}
\partial_{z} \left[ e^{-B(z)} \partial_{z} V \right]-p^2 e^{-B(z)}V=0, 
\label{eqmotion}
\end{equation}
where   $B(z)$ is  
\begin{equation}
B(z)=\log\left(\frac{z}{R}\right)+\phi(z)\,.
\label{B}
\end{equation}
The equation of motion (\ref{eqmotion}) has in general a discrete spectrum of normalizable solutions, $ V(p,z)=\Psi_n(z)$ that satisfy the boundary conditions $ \Psi_n(z=0)=0$ for $p^2=-m_{n}^{2}$. The eigenfunctions, $\Psi_n(z)$ are  normalized according to:
\begin{equation}
\int^{\infty}_{0}dz \ e^{-B(z)} \ \Psi_n(z)\Psi_m(z)=\delta_{mn} \,.
\label{Normalization}
\end{equation}
Decay constants are  proportional to the transition matrix of the vector meson in the $n$ state to the non hadronic state:  $  \langle 0 \vert \, J_\mu (0)  \,  \vert n \rangle = \epsilon_\mu f_n m_n $. 
They are calculated holographically  from the normalized solutions  $\Psi_n$ 
(see refs.   \cite{Karch:2006pv,Grigoryan:2007my})  as:
\begin{equation}
f_n=\frac{1}{g_{5} m_{n}}\lim\limits_{z \rightarrow 0} \left( e^{-B(z)}\Psi_n(z)\right) \,.
\label{decayconstant}
\end{equation}
In the case of the soft wall model the form of the dilaton is $\phi_{SW}(z)=k^2z^2$. The normalized solutions of equation (\ref{eqmotion}) vanishing at $z = 0$ have the form:
\begin{equation}
\Psi^{SW}_{n}=( k z) ^2 \ \sqrt{\frac{2}{n+1}}\ L^{1}_{n}( k z)  \,.
\label{SWsolution}
\end{equation}
The result for the decay constants in softwall model is
 \begin{equation}
 f^{SW}_n    \, =\,  
  k \sqrt{ 2}  / {\tilde g}_5  \,,
 \label{decayconstants}
\end{equation}  
\noindent where $ {\tilde g}_5^2 =  g_5^2 /R $. This means that in the soft wall model all the radial excitations of a  vector meson have the same decay constant. We show on table {\bf 1} the experimetal masses \cite{Agashe:2014kda} for charmonium states and the corresponding decay constants that decrease with  radial excitation level, contrasting with the result from the soft wall model. 
For the hard wall model \cite{Polchinski:2001tt,BoschiFilho:2002ta,BoschiFilho:2002vd}, relation  (\ref{decayconstant}) is also vaild. In this case  the decay constants increase with excitation level.

\begin{table}[h]
\centering
\begin{tabular}[c]{|c||c||c|}
\hline 
\multicolumn{3}{|c|}{  Charmonium data   } \\
\hline
State &  Mass  (MeV)     &   Decay constant (MeV) \\
\hline
$\,\,\,\, 1S \,\,\,\,$ & $ 3096.916\pm 0.011 $  & $ 416 \pm 5.3 $ \\
\hline
$\,\,\,\, 2S \,\,\,\,$ & $ 3686.109 \pm 0.012 $   & $ 296.1 \pm 2.5 $  \\
\hline 
$\,\,\,\,3S \,\,\,\,$ & $ 4039 \pm 1 $   & $ 187.1  \pm 7.6 $ \\ 
\hline
$ \,\,\,\, 4S  \,\,\,\,$ & $ 4421 \pm 4 $  & $ 160.8  \pm 9.7 $ \\
\hline
\end{tabular}   
\caption{Experimental masses and the corresponding decay constants for the Charmonium S-wave resonances.  }
\end{table}

In reference  \cite{Braga:2015jca}  it was shown that introducing a UV hard cut off in the soft wall model one can obtain decay constants decreasing with excitation level.  This model provided a consistent  picture for the thermal behavior of bottomonium states but not for charmonium. 
The dissociation of charmonium 1S state, the $ J / \Psi $, occurs at temperatures smaller than the critical one in this model.

In order to reproduce the  behavior of  $c \bar c$ vector mesons we  consider the following dilaton field:
\begin{equation}
\phi(z)=k^2z^2+\tanh\left(\frac{1}{Mz}-\frac{k}{ \sqrt{\Gamma}}\right)
\label{dilatonModi}
\end{equation} 
where  the parameter $k$ represents the charm quark mass, $\Gamma $ the string tension of the strong quark anti-quark interaction and $M$ is a mass scale associated with non hadronic decay.  The  values of the parameters that  describe  charmonium are:
\begin{equation}
  k = 1.2  \, {\rm GeV } ; \,\,   \sqrt{\Gamma } = 0.75  \, {\rm GeV } ; \,\, M=2.7  \, {\rm GeV }.
  \label{parameters}
  \end{equation}   
Note that $M$ is of the order of the mass change when a charmonium state undergoes a non hadronic decay. 
  
 The procedure to calculate masses and decay constants is to find the normalizable solutions $ \Psi_n(z)$ of eq. (\ref{eqmotion}), with the background of eq. (\ref{dilatonModi}),  that vanish at $z = 0$.  Then the numerical solutions are used in 
 eq. (\ref{decayconstants}).  We show  on table \textbf{2} the results for the mass and the decay constant using the holographic model describe above. One can observe  that the decay constants decrease with  radial excitation level, as in the experimental results of table 1. 
 In particular, the decay constant of the $1 S$ state, the $ J / \Psi$  is just $1.4 \% $ smaller than the one obtained from experimental data.  
 Defining the rms error for estimating $N$ quantities using a model with $N_p$ parameters as:
 \begin{equation}
 \delta_{rms} = \sqrt{ \frac{1}{(N - N_ p )}\sum_i^N  \left( \frac{\delta O_i}{O_i} \right)^ 2 }
 \label{error}
 \end{equation}
 \noindent where $O_i$ is the average experimental value and $\delta O_i$ is the deviation of the value obtained using the model, one finds $ \delta_{rms} =  20.7 \%$.

\begin{table}[h]
\centering
\begin{tabular}[c]{|c||c||c|}
\hline 
\multicolumn{3}{|c|}{  Holographic Results for Charmonium   } \\
\hline
 State &  Mass (MeV)     &   Decay constants (MeV) \\
\hline
$\,\,\,\, 1S \,\,\,\,$ & $ 2032.63 $  & $ 410.203 $ \\
\hline
$\,\,\,\, 2S \,\,\,\,$ & $ 3132.15 $   & $ 276.752 $  \\
\hline 
$\,\,\,\,3S \,\,\,\,$ & $ 3975.30  $   & $220.701$ \\ 
\hline
$ \,\,\,\, 4S  \,\,\,\,$ & $ 4640.60  $  & $ 190.846 $ \\
\hline
\end{tabular}   
\caption{Holographic masses and the corresponding decay constants for the Charmonium S-wave resonances.  }
\end{table}

\section{ Finite density plasma }

Once the zero temperature holographic description of charmonium was defined, let us extend the model to finite temperature and  density.  
Light vector mesons have been studied in the soft wall model at finite temperature in \cite{Mamani:2013ssa}. We consider the same action of eq. (\ref{vectorfieldaction}):
\begin{equation}
I \,=\, \int d^4x dz \, \sqrt{-g} \,\, e^{- \Phi (z)  } \, \left\{  - \frac{1}{4 g_5^2} F_{mn} F^{mn}
\,  \right\} \,\,, 
\label{vectorfieldaction2}
\end{equation}
but  with a charged black hole metric, as in 
\cite{Colangelo:2010pe,Colangelo:2011sr,Colangelo:2012jy} 
\begin{equation}
 ds^2 \,\,= \,\, \frac{R^2}{z^2}  \,  \Big(  -  f(z) dt^2 + \frac{dz^2}{f(z) }  + d\vec{x}\cdot d\vec{x}  \Big)   \,,
 \label{metric2}
\end{equation}
where:
\begin{equation}
f (z) = 1 - \frac{z^ 4}{z_h^4}-q^2z_{h}^{2}z^4+q^2z^6 \,.
\end{equation}
 The horizon position  $z_h$ is obtained from the condition $f(z_h)=0$. The temperature of the black hole is determined by the requirement of absence of  conical singularity at the horizon in the Euclidean version of the metric. One has: 
\begin{equation} 
T =  \frac{\vert  f'(z)\vert_{(z=z_h)}}{4 \pi  } = \frac{1}{\pi z_h}-\frac{  q^2z_h ^5}{2 \pi }\,.
\label{temp}
\end{equation}

The parameter $q$ is proportional to the black hole charge and related to the  quark chemical  potential $\mu$.  In the gauge theory side of the duality $\mu$ is a constant parameter that appears in the lagrangian multiplying the quark density 
$  \bar{\psi}\gamma^0 \psi \,$.  
So it plays the role of the source of correlators of  this operator. In the dual supergravity description the  field $V_0$ plays this role. 
So, one considers a particular solution for the vector field   $ V_m $ with only one non vanishing component: $  V_0 = A_0 (z) $  ($V_z =0, V_i = 0$). Assuming that the relation between $q $ and $\mu$ is the same as in the case of no background, that means $ \phi (z)  = 0$, the solution for the time component of the vector field is:
 $ A_{0}(z)=c - qz^2 $, where $c$ is a a constant. Imposing  $A_{0}(0) = \mu $ and $A_{0}(z_h)=0$ one finds: 
\begin{equation}
\mu = q z_h^2  \,.
\end{equation} 
The values of the dimensionless combination $ Q = qz_{h}^{3} $ must be in the interval
$0\leq Q \leq \sqrt{2}$. 

 The  finite temperature and density version of equation of motion (\ref{eqmotion}) reads
\begin{equation}
\partial_{z} \left[ f(z)e^{-B(z)} \partial_{z} V \right]-e^{-B(z)}\left( \frac{\omega^2}{f(z)}-|\vec{p}|^2   \right)V=0 \,,
\label{eqmotion2}
\end{equation}
where $V$ is  one of the spatial components  of the transverse gauge field. In the next section we show how to  can calculate the spectral function for charmonium.

\section{Spectral Function}
  
The charmonium spectral function can be evaluated using the membrane paradigm \cite{Iqbal:2008by} (see also \cite{Finazzo:2015tta}). Let us briefly review the membrane paradigm for  a vector field $V_\mu$, dual to the electric  current operator $J_{\mu}$. We assume a  black brane  metric of the  general  form:
\begin{equation}\label{metric3}
ds^{2}=-g_{tt}dt^2+g_{zz}dz^{2}+g_{x_1 x_1}dx^{2}_{1}+g_{x_2 x_2}dx^{2}_{2}+g_{x_3 x_3}dx^{2}_{3}\,.
\end{equation}
\noindent the metric components depend only on coordinate $z$ and the boundary is located at $z= 0$.  There is a horizon, where the time component vanishes:  $g_{tt}(z = z_{h})=0$. 
For the action of a vector field  we assume the general form
 \begin{equation}\label{Sigma2}
S=-\int d^{5}x \, \sqrt{-g}\frac{1}{4 g_5^2 \, h(z)}F^{m n}F_{m n}\,,
\end{equation}
where $h(z)$ represents a general background. The corresponding equation of motion is:
\begin{equation}\label{eqms}
\partial^{m} \left( \frac{\sqrt{-g}}{h(z)}  F_{m n } \right)=0 \,.
\end{equation}
Now we consider solutions for the vector field that are independent of coordinates  $x_1$ and $x_2$ and separate the equations of motion into a longitudinal channel involving fluctuations along $(t, x_3)$ and a transverse channel with fluctuations along the  spatial directions  $(x_1,x_2)$. 
The relevant components of eq. (\ref{eqms})  in this case read
\begin{equation}\label{Maxwell1}
-\partial_{z}j^{t}-\frac{\sqrt{-g}}{h}g^{tt}g^{x_3 x_3}\partial_{x_3}F_{x_3t}=0\,,
\end{equation}
\begin{equation}\label{Maxwell2}
-\partial_{z}j^{x_3}+\frac{\sqrt{-g}}{h}g^{tt}g^{x_3 x_3}\partial_{t}F_{x_3t}=0\,,
\end{equation}
\begin{equation}\label{Maxwellz}
\partial_{x_3}j^{x_3}+\partial_t j^{t}=0\,,
\end{equation}
\noindent where 
\begin{equation}\label{momentum}
j^{\mu}=-\frac{1}{h(z)}\sqrt{-g}F^{z \mu} \,.
\end{equation}
Using the Bianchi identity one finds:
\begin{equation}\label{Bianchi}
\partial_{z}F_{x_3t}-\frac{h(z)}{\sqrt{-g}}g_{zz}g_{x_3 x_3}\partial_{t}j^{z}-\frac{h(z)}{\sqrt{-g}}g_{tt}g_{x_3 x_3}\partial_{x_3}j^{t}=0\,.
\end{equation} 
A z-dependent (bulk) version of conductivity for the longitudinal channel can de defined as:
\begin{equation}\label{longitudinal}
\bar{\sigma}_{L}(\omega,\vec{p},z)=\frac{j^{x_3}(\omega,\vec{p},z)}{F_{x_3 t}(\omega,\vec{p},z)}\,.
\end{equation} 
For a plane wave solution with momentum  $p=(\omega,0,0,p_3)$ one finds, using  eqs.  (\ref{Maxwell1}),  (\ref{Maxwellz}) and (\ref{Bianchi}) 
\begin{equation}\label{Membrane}
\partial_{z}\bar{\sigma}_{L}=-i\omega\sqrt{\frac{g_{zz}}{g_{tt}}}\left[ \Sigma(z)-\frac{\bar{\sigma}_{L}^{2}}{\Sigma(z)} \left( 1-\frac{p_3^2}{\omega^2} \frac{g^{x_3 x_3}}{g^{tt}}          \right)    \right] \,,
\end{equation}
where 
\begin{equation}\label{Sigma}
\Sigma (z)=\frac{1}{h(z)}\sqrt{\frac{-g}{g_{zz}g_{tt}}}g^{x_3 x_3} \,.
\end{equation}
A similar procedure for the transverse channel leads to  \cite{Iqbal:2008by}:
\begin{equation}\label{Membrane2}
\partial_{z}\bar{\sigma}_{T}=i\omega\sqrt{\frac{g_{zz}}{g_{tt}}}\left[ \frac{\bar{\sigma}_{L}^{2}}{\Sigma(z)} - \Sigma(z)\left( 1-\frac{p_3^2}{\omega^2} \frac{g^{x_3 x_3}}{g^{tt}}          \right)    \right] \,.
\end{equation}

Here we want to consider only the case of zero momentum. So, we take $p_3^2 = 0$ and  both flow equations (\ref{Membrane}) and (\ref{Membrane2})  have the same form
\begin{equation}\label{MembraneF}
\partial_{z}\bar{\sigma}=i\omega\sqrt{\frac{g_{zz}}{g_{tt}}}\left[ \frac{\bar{\sigma}^{2}}{\Sigma(z)} - \Sigma(z)    \right] \,,
\end{equation}    
for $\bar{\sigma}= \bar{\sigma}_{T}=\bar{\sigma}_{L}$.

The AC  conductivity $\sigma$  is related to the Retarded Green function by the Kubo 
formula: $ \sigma(\omega)= i G_{R} ( \omega) / \omega\,$.  So, one defines: 
\begin{equation}\label{AC}
\sigma(\omega) =-\frac{G_{R}(\omega)}{i \omega}\equiv \bar{\sigma}(\omega,z = 0)\,.
\end{equation} 
Note that $Re$ $\sigma(\omega)=\rho(\omega)/\omega$, where $ \rho(\omega) \equiv - Im \ G_{R}(\omega)$ is the spectral function.

\begin{figure}[t]
\label{g67}
\begin{center}
\includegraphics[scale=0.7]{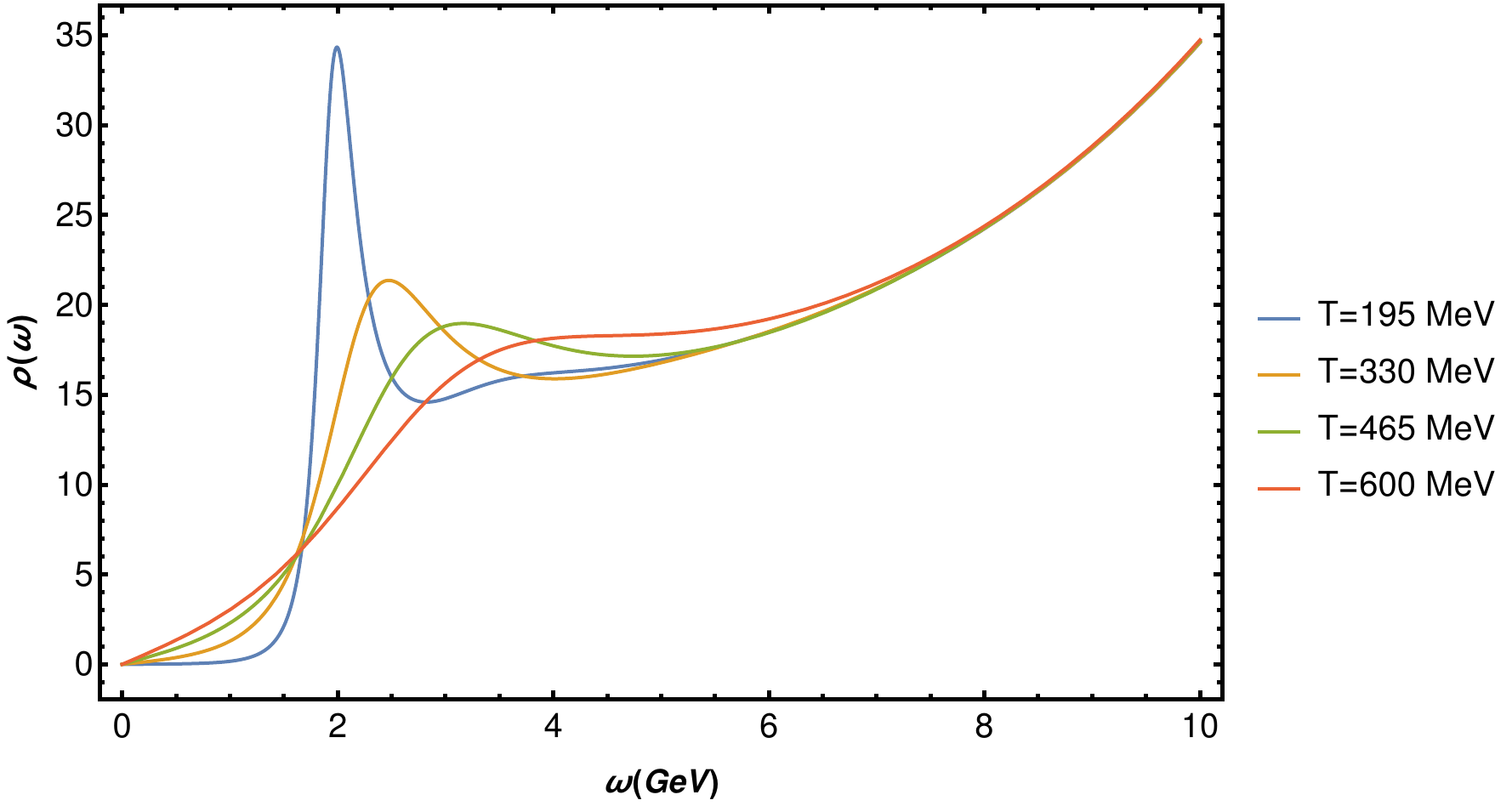}
\end{center}
\caption{ Spectral functions for $\mu=0$ at four representative  values of the temperature.}
\end{figure}

For the present model, described in the previous sections, we use the metric (\ref{metric2}) and $ h(z) = \exp \{k^2z^2 +\tanh\left(\frac{1}{Mz}-\frac{k}{ \sqrt{\Gamma }}\right)  \} $
 in the flow equation  (\ref{MembraneF}). One finds
\begin{equation}\label{MembraneF2}
\partial_{z}\bar{\sigma}(\omega,z)= \frac{i\omega}{f(z)\bar{\Sigma}(z)} \left[ \bar{\sigma}(\omega,z)^{2}-\bar{\Sigma}(z)^{2}  \right] \,,
\end{equation}      
with $\bar{\Sigma}(z)=\exp \{ -k^2z^{2} - \tanh\left(\frac{1}{Mz}-\frac{k}{ \sqrt{\Gamma }}\right) \}/z$. Requiring regularity at the horizon, one obtains the following condition: $ \bar{\sigma}(\omega,z_h)=\bar{\Sigma}(z_{h})\,.$ The spectral function is obtained from:  
\begin{equation}\label{spectralfunction}
\rho(\omega)\equiv - Im \ G_{R}(\omega)= \omega  Re \  \bar{\sigma}(\omega, 0)\,.
\end{equation}

\section{Results and Discussion}

The strategy  used to obtain the spectral function for charmonium is to evaluate numerically equation (\ref{MembraneF2})  with the boundary conditions described in the previous section. The parameters  used are the ones that provide the best fit in the  zero temperature case of section II namely a mass scale  $ M=2.7 $ GeV associated with the decay process, a quark mass of $k=1.2 $  GeV and $  \sqrt{\Gamma } =0.75$  GeV, where $ \Gamma$ represents the string tension. 
 
\begin{figure}[h]
\label{g67}
\begin{center}
\includegraphics[scale=0.3]{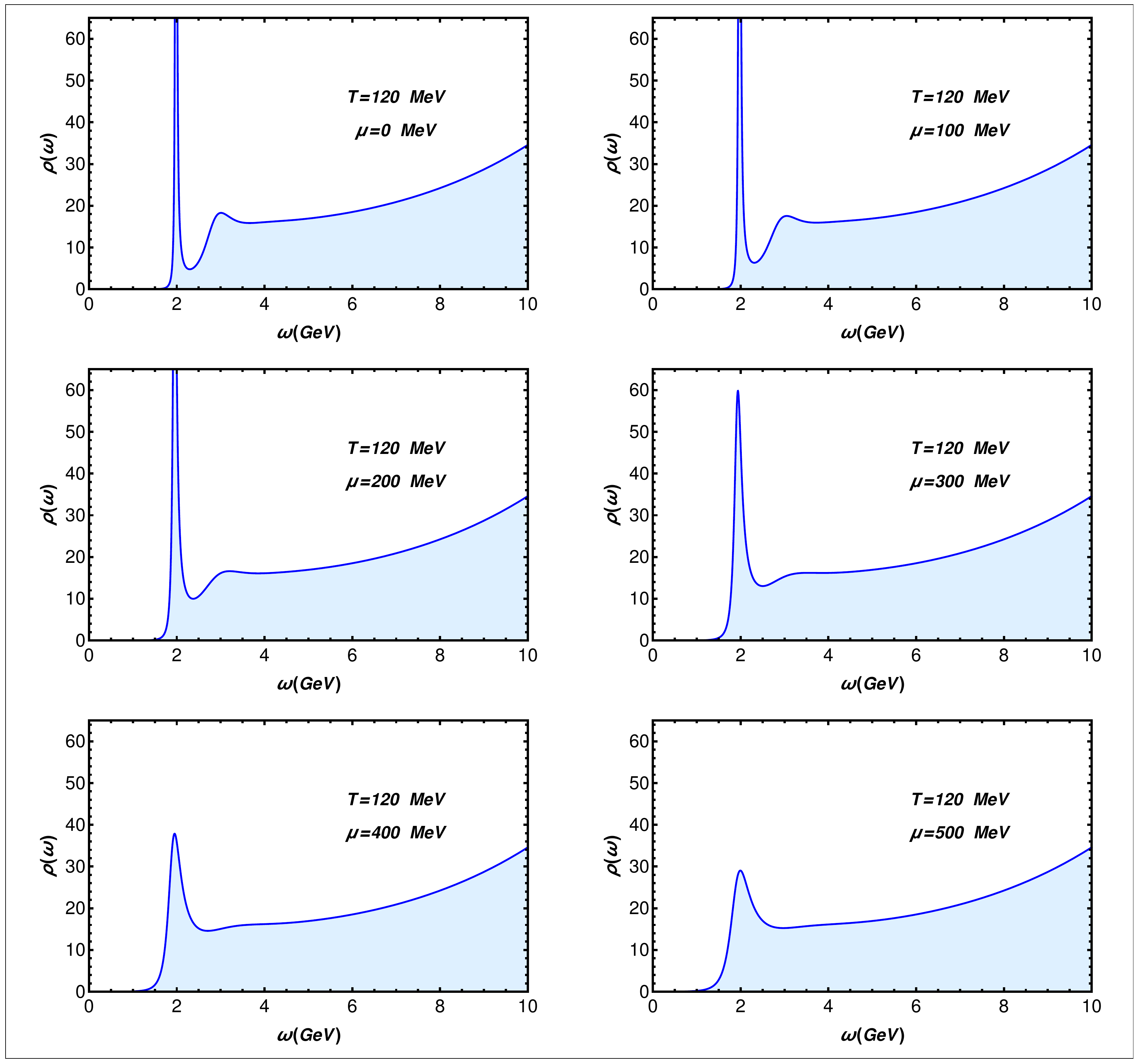}
\end{center}
\caption{Spectral functions for $T=$ 120 MeV at 6 representative  values of $\mu$  }
\end{figure}

We show in figure {\bf 1} the spectral function for $c \bar c$ vector mesons at four different temperatures, with zero chemical potential,  that illustrate  clearly the thermal  dissociation process. At $T= 195 $ MeV there is a clear peak corresponding to the $ 1 S$  state, the $J /\Psi$. Increasing  the temperature the peak decreases  showing the thermal dissociation  process. 

For the case of non vanishing  chemical potential we start the analysis with  $T= 120 $ MeV (supercooled phase) for six  different values of  chemical potential from $ \mu = 0 $ to $ \mu = 500 $ MeV, as shown in figure {\bf 2 }.  One can see  on the first  panel the peak corresponding to the $1 S$ state and a very small peak corresponding to the $2S$ state that is already dissociated in the medium. Then, raising the value of $\mu $ the $ 1 S $ peak develops a larger width and a smaller height, representing the dissociation as the effect of the density of the medium.

\begin{figure}[h]
\label{g67}
\begin{center}
\includegraphics[scale=0.3]{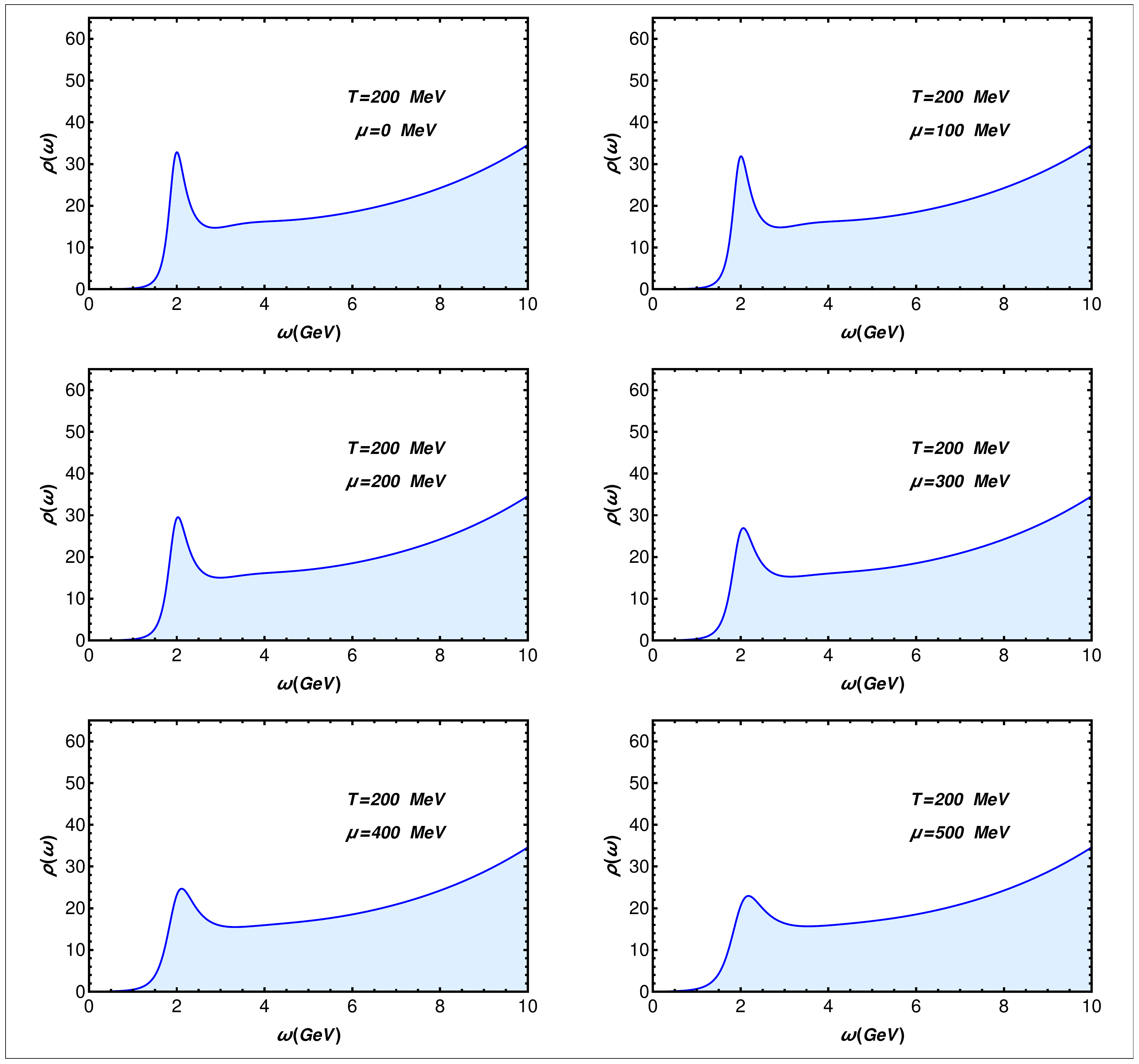}
\end{center}
\caption{Spectral functions for $T=$ 200 MeV at 6 representative  values of $\mu$  }
\end{figure}

Then in figure {\bf 3 } we show the case of $ T= 200 $ MeV. At zero chemical potential the second quasi-particle peak associated with the $2 S$ state   completely  disappeared and the $1 S$ state is partially dissociated in the medium. Increasing $\mu$ one sees that this state strongly increases the degree of  dissociation  in the medium.

 A compilation of results for quarkonium dissociation temperatures using lattice QCD and potential models  is presented in \cite{Adare:2014hje}.  The dissociation at zero chemical potential happens for the ration $ T/T_c$ in the range of 1.5 to 3.0. The critical temperature in the soft wall model is $ T_c \sim 190$ MeV and we assume that this critical temperature holds here also. So, the dissociation temperatures found here for the zero density case are consistent with the lattice predictions.  On the other hand, figures {\bf 2 } and {\bf 3} show that the dissociation by the effect of the density of the medium occurs when the quark chemical potential $\mu $ is of the order of 500 MeV. 

It is important to mention that some interesting studies of heavy mesons using holography appeared before in, for example,  refs. \cite{Hong:2003jm,Kim:2007rt,Fujita:2009wc,Noronha:2009da,Fujita:2009ca,Grigoryan:2010pj,Branz:2010ub,Gutsche:2012ez,
Afonin:2013npa,Hashimoto:2014jua,Liu:2016iqo,Liu:2016urz}.   However a  consistent picture for  the dissociation of  $J /\Psi$ in a plasma with finite temperature and density was missing  in the literature. 

As a final remark, we want to stress the fact that we constructed a phenomenological model with the purpose of describing the thermal behavior of charmonium states in a finite medium. 
Presently,  it is not possible to investigate the dissociation of quarkonium states in such a medium using lattice QCD. The model does not come from QCD. It comes from a particular way of breaking conformal invariance in the AdS/CFT duality, represented by the introduction of the background of eq. (\ref{dilatonModi}).  However it provides a picture for the spectral function of charmonium quasi-particle states that is qualitatively consistent and could be confronted in the future with other approaches.

 \noindent {\bf Acknowledgments:}   N.B. is partially supported by CNPq (Brazil) and L. F.  is supported by CAPES (Brazil) and A.V.  supported by FONDECYT (Chile) under Grant No. 1141280.

 \end{document}